\documentclass[10pt]{article}
\usepackage[sort&compress,numbers,super]{natbib}
\usepackage{amssymb}
\usepackage[version=3]{mhchem}
\usepackage[paperwidth=210mm,paperheight=297mm,total={16.0cm,24.7cm}, top=1.5cm, includeheadfoot, centering, footskip=\footskip+4mm ]{geometry}
\usepackage{mathptmx}
\usepackage{graphicx} 
\usepackage{float}
\usepackage[english]{babel}
\usepackage{charter}
\usepackage[compact]{titlesec}
\usepackage{hyperref}
\usepackage{subfig}      
\usepackage{xcolor}
\usepackage{nameref}
\usepackage[font=small,labelfont=bf]{caption}
\usepackage[normalem]{ulem}

\begin{document}
\begin{center}
\LARGE{\textbf{ArGSLab: A tool for analyzing experimental or simulated particle networks}}
\end{center}

\begin{center}
\large{Jasper N. Immink$^{\dagger,^\parallel,\ast}$, J. J. Erik Maris$^\ddagger$, Ronja F. Capellmann$^\dagger$, Stefan U. Egelhaaf$^\dagger$, Peter Schurtenberger$^{\parallel,\perp}$, Joakim Stenhammar$^\parallel$}\\
\normalsize{\textit{$^\dagger$ Condensed Matter Physics Laboratory, Heinrich-Heine-University Düsseldorf, Düsseldorf, Germany.\\ $^\parallel$ Division of Physical Chemistry, Lund University, Lund, Sweden.\\ $^\ddagger$ Inorganic Chemistry and Catalysis Group, Utrecht University, Utrecht, the Netherlands.\\  $^\perp$ Lund Institute of advanced Neutron and X-ray Science (LINXS), Lund University, Lund, Sweden.\\ $^\ast$ Tel: +49-211 8114327; E-mail: immink@hhu.de}}
\end{center}

\textbf{Microscopy and particle-based simulations are both powerful techniques to study aggregated particulate matter such as colloidal gels. The data provided by these techniques often contains information on a wide array of length scales, but structural analysis methods typically focus on the local particle arrangement, even though the data also contains information about the particle network on the mesoscopic length scale. In this paper, we present a MATLAB software package for quantifying mesoscopic network structures in colloidal samples. ArGSLab (Arrested and Gelated Structures Laboratory) extracts a network backbone from the input data, which is in turn transformed into a set of nodes and links for graph theory-based analysis. The routines can process both image stacks from microscopy as well as explicit coordinate data, and thus allows quantitative comparison between simulations and experiments. ArGSLab furthermore enables the accurate analysis of microscopy data where, \emph{e.g.}, an extended point spread function prohibits the resolution of individual particles. We demonstrate the resulting output for example datasets from both microscopy and simulation of colloidal gels, in order to showcase the capability of ArGSLab to quantitatively analyze data from various sources. The freely available software package can be used either with a provided graphical user interface or directly as a MATLAB script.}\\


\section{Introduction}
Colloidal dispersions exist in many shapes and forms, and understanding their behavior is relevant to biological,\cite{Don2003} ecological,\cite{Jackson2009} and industrial\cite{Alexopoulos2007} processes. The structures observed in such dispersions include for example amorphous colloidal glasses\cite{Weeks2012} and arrested colloidal gel networks.\cite{Pusey1995,Sandkuhler2004} The macroscopic behavior of colloidal dispersions is in large part determined by this spatial arrangement, and material properties such as the thermal conductivity and rheology vary strongly with the micro- and mesoscopic organization of the particles.\cite{Aksay1990,Lin2003,Egelhaaf2009,Burgi2010,Schurtenberger2013a} Thus, determining this spatial organization and understanding its link to the macroscopic behavior is a major point of interest within soft matter science.\cite{Dickinson2013,Wagner2017,Furst2019} For example, knowledge about the local structure of colloidal particles in gel networks\cite{Tanaka2008,Sprakel2017,Hermes2019} has led to the development of models that couple the network structure to macroscopic parameters such as the storage modulus.\cite{Morbidelli2001} Such models often require structural data that describe the gel network over a broad range of length scales, and microscopy and numerical simulations both offer powerful tools for probing the formation kinetics and structural properties of particle gels on this broad range of length scales.\cite{Solomon2003,Bartlett2005,Helgeson2015,Vermant2020,Walstra1995,Cates2002,Sandy2017,Zia2018} These methods are often combined with computational analyses that allow for the extraction of structural parameters,\cite{Grier1996,Weitz2001b} such as pair correlation functions, structure factors, and coordination numbers, that are particularly useful for describing the local network structure.\cite{Vermant2011,Solomon2012,Cohen2018,Simonsen2019} \\ 
Applying these analyses to experimental and computational datasets,\cite{Sciortino2008,Schmiedeberg2016,Furst2019} valuable insight regarding the local structure of such particle networks can be gained. However, such datasets also contain structural information on longer length scales, although a limited number of tools exist that can analyze the datasets reliably and efficiently.\cite{Dolors2007,Eiser2016} Existing analysis tools primarily focus on quantifying the material thickness and porosity of the network.
These analyses however require well-resolved individual network segments, which is not always available when analyzing microscopy data due to limitations in image resolution, especially with small particles.\cite{Egelhaaf2008,Xiaowei2009,Stenhammar2020}\\ 
In this Article, we will present a versatile computational method called ArGSLab, Arrested and Gelated Structures Laboratory, developed as a MATLAB package providing a set of tools to quantitatively analyze network structures in particle aggregates such as colloidal gels. ArGSLab performs a mesoscopic structural analysis that uses experimental or computational data to reconstruct three-dimensional gel networks\cite{Tanaka2019} as graphs of nodes and links.\cite{Stenhammar2020} ArGSLab furthermore enables quantification of the network structure using mesoscopic structural parameters such as node densities, branching densities and branch length statistics. As input, ArGSLab handles both image stacks from microscopy and sets of particle coordinates from either experiment or simulation. Crucially, the software does not require microscopy data containing well-resolved particle positions, enabling the analysis of structures formed by particles down to the same size range as the point-spread function. The software then converts the input data into single voxel-thin backbone structures, using routines based on established algorithms.\cite{Groen1980,Chu1994,Fratzl2017} The generated network structures are then transformed into a set of nodes and links for further statistical, graph theory-based analysis and visualization. ArGSLab can be accessed either via the provided graphical user interface (GUI) or directly through the MATLAB script. \\
In the following, we will first summarize the main features of ArGSLab and briefly discuss the overall workflow. Then, we will discuss the algorithms in some detail, and thereafter discuss the different output options. Finally, we will illustrate the capabilities of the software using three different data sets: (i) a set of microscopy images of colloidal gels with well-resolved particles, (ii) similar data but with poorly resolved individual particles, and (iii) coordinate sets from Brownian dynamics simulations. 

\section{Software description}

\begin{figure}[ht]
	\centering
	\includegraphics[width=\linewidth]{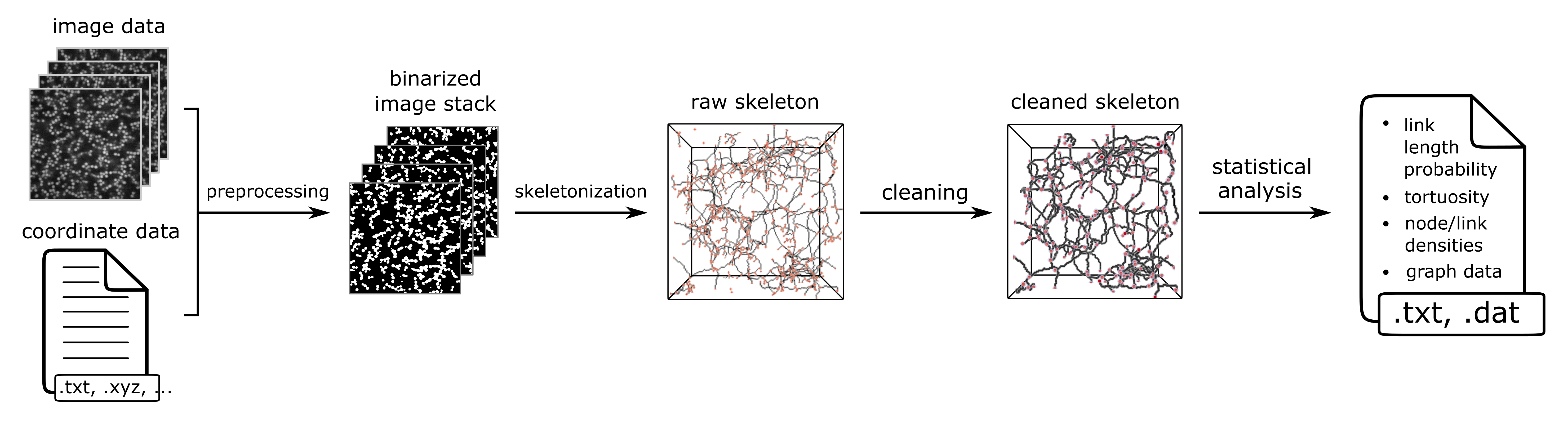}
	\caption{Overview of the ArGSLab workflow. As input, the software uses either image stacks or particle coordinate sets, which are preprocessed into binarized images and then transformed into ``skeletons''. These go through a cleaning step, and are then transformed into a set of nodes and links for quantitative analysis. ArGSLab provides a histogram of the link lengths, the density of nodes and links in the system, and the tortuosity of the network.}
	\label{Pipeline}
\end{figure}

An overview of the general ArGSLab workflow is shown in Fig.~\ref{Pipeline}. The four major stages of the analysis are: (i) preprocessing, (ii) skeletonization, (iii) cleaning and (iv) quantitative analysis. In the preprocessing stage, coordinate sets are first transformed into image stacks, which are then successively Gaussian blurred, binarized and morphologically closed. Small unconnected structures are then removed, and small holes are filled. The skeletonization stage consists of the iterative erosion of voxels that belong to the material network, until a single-voxel thick backbone skeleton remains. This is then transformed into a graph of nodes and links. The cleaning stage removes artefacts from the skeletonization process, and the fourth and final stage contains a thorough analysis of the graph, producing statistics such as link length distribution, tortuosity and node density. In the following, we will describe these features.\\

\subsection{Algorithms}
\label{secalgo}
As input, the package accepts either .tiff image stacks from confocal microscopy or 3D particle coordinate sets from, \textit{e.g.}, computer simulations. Coordinate diagrams are first projected onto a .tiff image stack using user-defined voxel dimensions, with an optional convolution using a default or user-defined point spread function. The user-provided or generated image stacks are then treated with a Gaussian blur and binarized (\emph{i.e.}, transformed into images containing only black or white pixels). The purpose of treating images with a Gaussian blur is to reduce noise in the final skeleton, predominantly in the form of unrealistically small side chains, a known artefact of this type of skeletonization.\cite{Chu1994,Tougne2015} The binarization step is performed with a threshold individually calculated for each image slice. Determination of the threshold can be done either manually or using automated routines, where the automatic determination uses Otsu's method,\cite{Otsu1979} whereas the manual method determines, for each slice, the intensities $I_{10}$ and $I_{90}$, which are the pixel intensity values below which 10\% and 90\% of individual pixel intensities fall, respectively. A user-defined value $V_u$ between 0 and 1 then defines the threshold intensity $I_T$, with
\begin{equation}
I_T = I_{10} + V_u(I_{90}-I_{10}).
\end{equation}
The threshold determination is performed on each slice individually in order to compensate for differences in signal intensities between image slices: this is helpful when investigating microscopy data where a scattering length density mismatch leads to the average fluorescence intensity being dependent on the focal depth in the sample, or where fluorescence bleaching effects play a role.\cite{Egelhaaf2008} Images are then treated with a morphological closing step, consisting of subsequent dilation and erosion steps~\cite{Dougherty1992} performed with a user-defined ellipsoidal structural element. This procedure removes noise and effectively fills small gaps between particles, ensuring contiguity where necessary.\cite{Devaux2011} Structures are connected should their binarized shapes be closer than twice the structural element radius. Subsequently, unconnected clusters smaller than a threshold fraction are removed, with larger unconnected clusters being analyzed as separate network structures. Finally, holes in the structure that are enclosed on all sides in 3D and are smaller than a user-defined threshold value are detected and filled in, reducing unphysical holes in the skeleton. The sizes of the Gaussian blur kernel, the morphological closing element and the minimal cluster size take either default or user-defined values.\\ 
Subsequently, the package performs the skeletonization, using algorithms adapted from \citet{Fratzl2017} This process consecutively removes material voxels bordering non-material voxels, with the condition that they are not end points, non-Euler invariant points, and are Euler simple points. Removal of a voxel considers the 3$\times$3 voxel cube $D_3(i)$ surrounding voxel $i$, \textit{i.e.} a cube of 27 voxels with voxel $i$ at the center. An end point is defined as the case where $D_3(i)$ contains in total two material voxels, of which one is $i$. Euler invariant points are defined as voxels that, upon removal, do not lead to a change of the Euler characteristic of the total structure.\cite{Pudney1998,Hansen2010,Baja2016} We define an Euler simple point as a voxel $i$ whose removal neither changes the connectivity nor the Euler characteristic of $D_3(i)$;\cite{Pudney1998,Omar2013} a more detailed discussion of these definitions can be found in \citet{Chu1994} The transformation of a skeleton into a graph of nodes and links is then performed by assuming that (i) each voxel $i$ whose surrounding $D_3(i)$ contains exactly 2 material voxels is an end node, (ii) each voxel $i$ whose $D_3(i)$ contains exactly 3 material voxels belongs to a link and (iii) each voxel $i$ whose $D_3(i)$ contains more than 3 material voxels is a branching node.\\
While the skeletonization process robustly extracts the relevant backbone structure, several known unrealistic features can arise, often due to noisy or irregular data.\cite{Chu1994} The subsequent cleaning steps remove such unwanted features, and are performed iteratively using the following procedures:
\begin{enumerate}
\item Unphysically small side chains, arising from irregular structure surfaces,\cite{Chu1994,Chen2018} are excluded by removing all side chains smaller than a certain threshold value, by default set to 1.5$\sigma$ with $\sigma$ the particle diameter. 
\item The removal of \textit{canal nodes}, \textit{i.e.} nodes that are connected to exactly two links. Such nodes and their connecting links are effectively one single link and therefore merged.
\item Nodes that are in close proximity are collected into a single node, as dense sections of gels with irregular edges often lead to complex and unphysical skeleton sections with many interconnected nodes and links. A set of nodes to be collected contains all nodes that are closer than a threshold value to another node in the same set. The threshold value is by default set to 0.9$\sigma$.
\end{enumerate}
After these procedures, a new skeleton is formed from the cleaned graph.\\

\begin{figure}[ht]
	\centering
	\includegraphics[width=.5\linewidth]{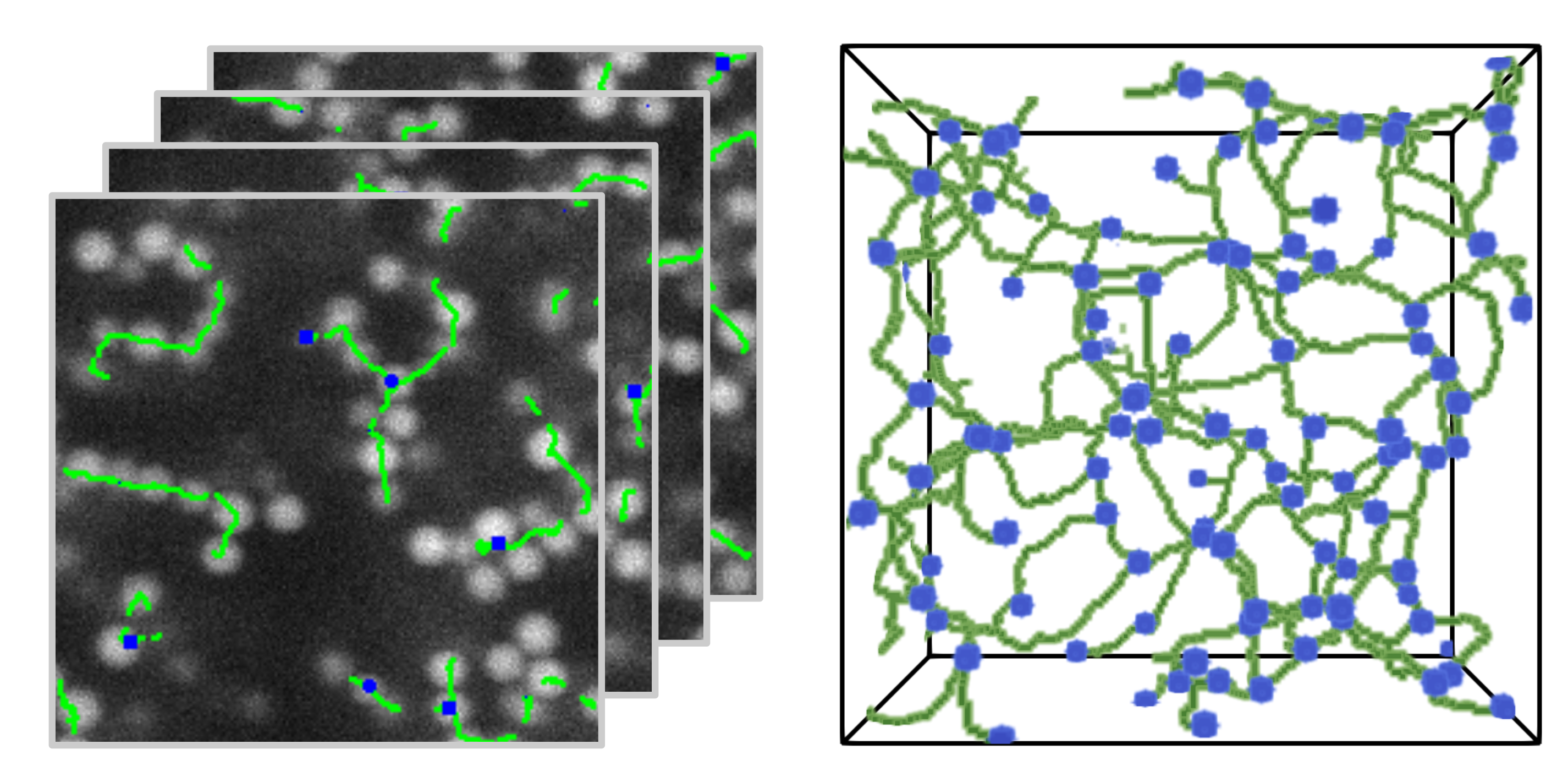}
	\caption{Visualized output. A) slices of the original image .tiff stack overlaid with the skeleton (green) and branching nodes (blue). Since these are 2D slices of a 3D image, interrupted skeleton strands continue in adjacent $z$-slices. B) 3D rendering of the same skeleton.}
	\label{vis_res}
\end{figure}

\subsection{Visualization}
\label{secvis}
ArGSLab provides output that allows the user to visualize and quantify the analyzed network structures. The software also provides output that helps optimizing the parameters of the skeletonization process. The standard visualizations are: (i) the original image stack overlaid with the calculated skeleton and branching nodes, as shown in Fig.~\ref{vis_res}A (data from \citet{Schmiedeberg2016}), and (ii) a fully rotatable 3D rendering of the skeleton backbone, as shown in Fig.~\ref{vis_res}B.\cite{Law2005,Yeh2020}\\
\begin{figure}[t]
	\centering
	\includegraphics[width=\linewidth]{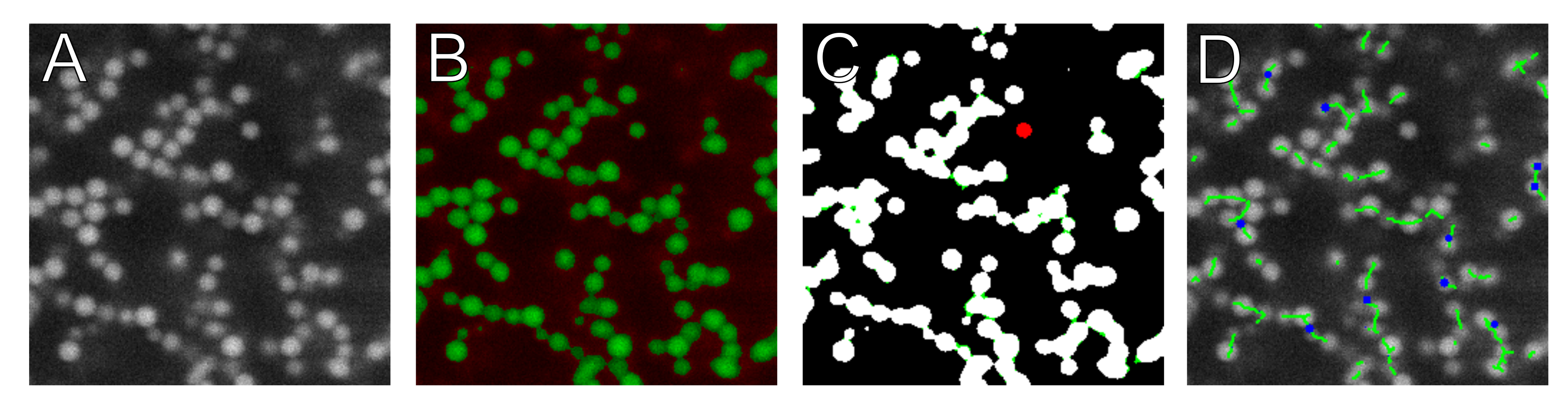}
	\caption{Slices from 3D input and a selection of ArGSLab's output image stacks. A) Raw input image. B) Raw input image overlaid with a green and red color filter, produced in the binarization step. Green indicates voxels determined to contain material (material voxels), and red to contain no material (non-material voxels). Black voxels had little to no intensity in the original image. C) Preprocessing effects after binarization: white sections are material voxels, unaltered by the preprocessing, green sections are material voxels added by preprocessing, and red sections are previous material voxels that have been removed. D) Overlay of the raw data with the final skeleton and nodes.}
	\label{Optimization}
\end{figure}As described in Section~\ref{secalgo}, several parameters and thresholds can be changed from their default values to optimize the output. The package provides several types of visualization to help optimize those parameters, a selection of which are shown in Fig.~\ref{Optimization}. Starting from a raw image stack (Fig.~\ref{Optimization}A), a new image stack is output after the binarization step (Fig.~\ref{Optimization}B). This is intended to help setting the binarization threshold, should the automatic method be insufficient. Fig.~\ref{Optimization}C shows the effects of further preprocessing after binarization, \textit{i.e.} the morphological closing, hole filling, and the removal of smaller unconnected structures. The final results are shown in Fig.~\ref{Optimization}D, showing an overlay of the raw image with the final skeleton.\\

\subsection{Statistical analysis}
From the graph of branching nodes and connecting links, a number of statistical quantities can be extracted, such as the number and density of nodes and links. The total number of nodes and links $N_N$ and $N_L$ are calculated together with the average number of links per node $\frac{N_L}{N_N}$. The density of nodes and links $\rho_N$ and $\rho_L$ respectively are given in units of ${\mu}\text{m}^{-3}$ and $\sigma^{-3}$ with $\sigma$ the particle diameter. Furthermore, ArGSLab provides a histogram showing the distribution of link lengths in the network, presenting the normalized link count $\tilde{N}(\Lambda)$ as a function of link length $\Lambda$, which we define as the length of a link between two nodes as measured along the link path. $\tilde{N}(\Lambda)$ is related to the fractal dimension $d_f$, a parameter often used for characterizing aggregated systems:\cite{Aksay1990,Morbidelli2003,Morbidelli2016} a propensity for long links and low $d_f$ are both characteristic of open aggregates. Another output measure is the tortuosity $\xi$, which is defined as the average trajectory length following a backbone strand $\lambda$, divided by the calculated Euclidean distance $\lambda_{Euc}$ between their start and end points A and B:
\begin{equation}
\xi = \left\langle \frac{\lambda(\text{A,B})}{\lambda_{Euc}(\text{A,B})} \right\rangle,
\end{equation}
where A and B are the start and end points of paths along the backbone intersecting respectively the upper and the lower box face in either Cartesian direction, and the angular brackets denote averaging over all such paths for a given configuration. $\xi$ provides a measure of the strand erraticity,\cite{Gladden1992,Liu2006,Harvey2018,Stenhammar2020} and is often used in catalysis,\cite{Hu2017} blood vessel analysis,\cite{Tomaszewski2018} or satellite geography.\cite{Harvey2018} All quantitative analyses are corrected for box edge effects. Furthermore, additional analyses beyond the ones described here are facilitated by ArGSLab as the skeleton backbone and the node and link lists can be exported as .mat or .txt files, after which analyses tailored to the specific system can be applied.\\

\begin{figure}
	\centering
	\includegraphics[width=.6\linewidth]{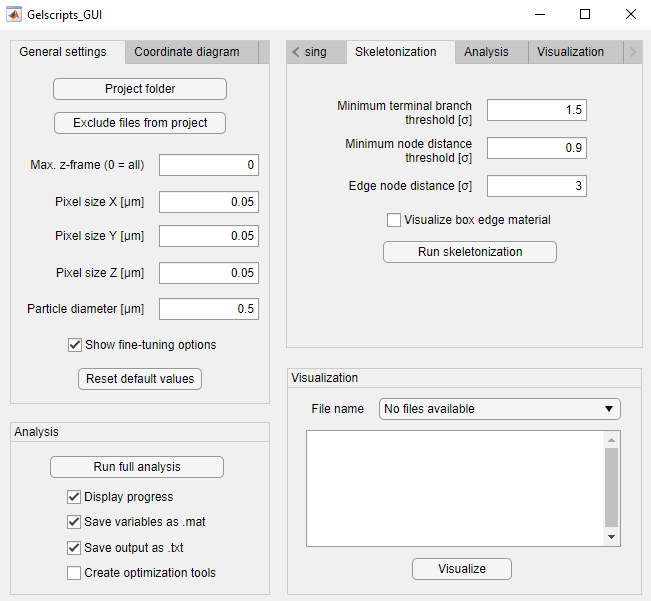}
	\caption{Screenshot of the GUI, with fine-tuning tools enabled.}
	\label{GUI}
\end{figure}

\subsection{Graphical User Interface}
The package can be accessed either using the provided MATLAB wrapper code, or using the graphical user interface (GUI), shown in Fig.~\ref{GUI}. The GUI gives intuitive access to the full capabilities of the code and requires only a few inputs: the folder path containing the input data, the number of $z$-slices to be analyzed in each file (in case not all slices should be analyzed), the voxel dimensions, and the average particle diameter. Ticking the ``fine-tuning options'' box reveals several further options that allow for optimization of output quality and can typically be left at their default values. These are described in-depth in the GUI tooltip menus and the package manual. The GUI is designed to be an intuitive tool for using ArGSLab, allowing users to run the full analysis without loss of functionality as compared to the wrapper script.\\

\section{Examples}
\begin{figure}
	\centering
	\includegraphics[width=\linewidth]{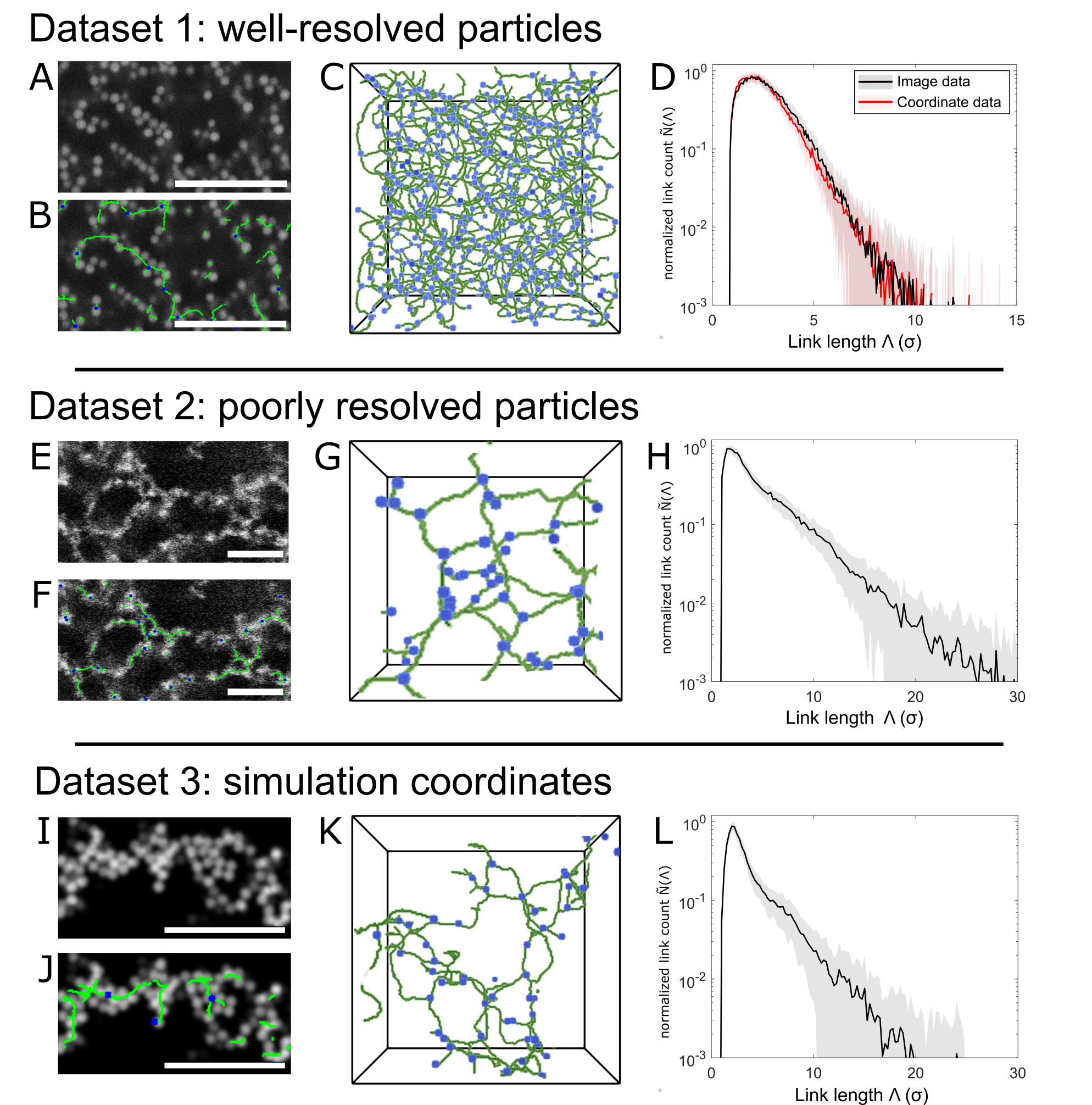}
	\caption{Analysed data from the three example datasets: A-D) Dataset 1, corresponding to dataset C5 in Ref.~\citenum{Schmiedeberg2016}. E-H) Dataset 2, containing particle gels formed by attractive microgel particles, synthesized as in Ref.~\citenum{Schurtenberger2020} I-L) Brownian dynamics simulation data of particles with patchy attractions, from Ref.~\citenum{Stenhammar2020}. Panels (A,E,I) show image sections from the raw data, panels (B,F,J) overlays of the same image and the final skeleton, panels (C,G,K) 3D renderings of the full skeleton (green) and  branching nodes (blue), and panels (D,H,L) the normalized link length distribution $\tilde{N}(\Lambda)$. The two datasets in panel (D) corresponds to $\tilde{N}(\Lambda)$ obtained using either the raw image data, or from a particle coordinate set extracted from the same data, demonstrating the equivalence of the two approaches. All 3D skeleton renderings correspond to a volume of approximately 1.7 $\times 10^4 \sigma^3$, and the scale bars represent 10$\sigma$.}
	\label{HSExample}
\end{figure}

In the following we will illustrate the use of ArGSLab by showing the processing of three example datasets. Dataset 1 consists of microscopy images of a set of colloidal gels made from PMMA spheres with an added depletant that causes attractive forces leading to gel formation. The gels are imaged using confocal laser scanning microscopy (CLSM) and the large particles allow for individual particles to be well-resolved.\cite{Schmiedeberg2016} The results from analyzing this dataset are compared to the corresponding results from coordinate sets of the particle centers, extracted using routines following Crocker and Grier.\cite{Grier1996} Dataset 2 contains CLSM images of gels of smaller, pNIPAm microgel particles (as described in Section~\ref{secmatmet}) that cannot be individually resolved. Dataset 3 is a set of particle coordinates obtained from Brownian dynamics simulations of colloidal gels formed from colloidal particles with attractive patches icosahedrally arranged on the particle surface.\cite{Stenhammar2020} Further details about the preparation can be found in Section~\ref{secmatmet}, and the results are summarized in Fig.~\ref{HSExample} and Table~\ref{HSExampleTab}.\\
Figs.~\ref{HSExample}A-D show the results for Dataset 1. The quantitative analyses in Fig.~\ref{HSExample}D and Table~\ref{HSExampleTab} reflect the similarity between experimental skeletons and the ones obtained from the corresponding coordinate data. The small discrepancies between the results can be explained as follows: the coordinate extraction routines identify the vast majority of particles, but atypical particles, due to synthetic or imaging anomalies, are sometimes not picked up. While this only leads to negligible errors in tracking experiments, ArGSLab's structural analysis identifies this as broken links. Such effects can also arise from the determination of a particle center slightly off from its real particle center due to experimental noise. ArGSLab can partially accommodate such anomalies by choosing larger structural element sizes. \\
The results from Dataset 2 (Figs.~\ref{HSExample}E-H and Table~\ref{HSExampleTab}) clearly demonstrate the software's ability of extracting quantitative data from structures constructed from non-resolvable particles. Nonetheless, the lower data quality often leads to the need for a more careful selection of optimization parameters. Skeletonization of data with low contrast, low signal-to-noise ratio, or poor $z$-resolution is known to lead to skeletons with small, unphysical side chains, or with a multitude of nodes in dense cluster-like sections.\cite{Chu1994} Tuning the parameters of the cleaning routines can alleviate these types of noise: larger Gaussian blurring kernels or morphological closing structural elements often cause dense clusters to be correctly interpreted as closed entities. Furthermore, the node collection threshold in the second cleaning step can be increased, collecting these virtual nodes into a single node. Furthermore, the terminal branch length threshold (third cleaning step in Section~\ref{secalgo}) can be raised to remove unphysical side chains. Changing these parameters can affect the number of nodes and links found, and should be kept constant between datasets for full comparability. In all our examples we have used the default parameters, which provides optimal comparability between different experiments.\\
One important consideration when setting the analysis parameters is that determining whether two particles are to be considered in contact or not is a non-trivial question\cite{Makse2007,Egelhaaf2011a,Hsiao2020} which will affect the final skeleton. Whether ArGSLab considers two segments in contact or not is implicitly governed by the binarization and the morphological closing steps. In brief, two gel segments will be considered in contact if their material voxels after binarization are closer together than the diameter of the morphological closing element. This choice is rationalized by considering that colloidal gels require a significant attractive potential, and intersegment gaps are often metastable, while noisy images often lead to apparent gaps. Optimal skeleton output therefore requires a careful tuning of the morphological closing element size and the binarization threshold, with the preselected standard parameters as good starting points.\\
The results from Dataset 3 (Figs.~\ref{HSExample}I-L and Table~\ref{HSExampleTab}) further demonstrate that ArGSLab offers a direct comparison between experimental systems and model systems studied by simulations. The large difference in the node and link densities $\rho_N$ and $\rho_L$ between Dataset 1 and Datasets 2 and 3 results from the much higher volume fraction in the former system ($\phi = 0.2$ compared to $\phi = 0.05$) that leads to much higher node and link densities in Dataset 1. The difference in $\phi$ also results in a significantly lower $\tilde{N}(\Lambda)$ at high $\Lambda$ for Dataset 1, and the tortuosity $\xi$ is also lowered by the higher $\rho_N$. It is also apparent that the gels in Datasets 2 and 3 have a stronger propensity for having end nodes rather than branching nodes from the lower $N_L/N_N$ ratios in Table~\ref{HSExampleTab}.\\

\begin{table*}
\centering
\caption{Values from statistical analysis of the example datasets in Fig.~\ref{HSExample}.}
\label{HSExampleTab}
\begin{tabular}{ l | l | l | l | l }
Type & $\rho_N$ [$\sigma^{-3}$] & $\rho_L$ [$\sigma^{-3}$] & $N_L/N_N$ & $\xi$\\ 
\hline
\hline
Dataset 1 (images) & $7.86 \times 10^{-2}$ & $1.34 \times 10^{-1}$ & $1.70$ & $2.03 \pm 0.14$ \\
Dataset 1 (coordinates) & $8.36 \times 10^{-2}$ & $1.44 \times 10^{-1}$ & $1.72$ & $2.05 \pm 0.09$ \\
\hline
Dataset 2 & $7.18 \times 10^{-3}$ & $8.67 \times 10^{-3}$ & $1.22$ & $2.23 \pm 0.12$ \\
\hline
Dataset 3 & $9.10 \times 10^{-3}$ & $1.02 \times 10^{-2}$ & $1.13$ & $2.25 \pm 0.13$ \\
\end{tabular}
\end{table*}

\section{Conclusions and Perspectives}
In this Paper, we have demonstrated the capabilities of ArGSLab, a tool for analyzing networks of aggregated particulate matter, such as colloidal gels. We have shown that ArGSLab is able to analyze both microscopy and simulation data, and allows for direct comparison between different datasets. The software provides a quantitative analysis of the mesoscopic network structure, which is more difficult to analyze than the local particle arrangements. Importantly, ArGSLab also provides a method to quantify systems where individual particles are not resolvable and the local structure thus cannot easily be probed. The graph theory approach that our package applies enables comprehensive analyses of aggregated structures, and it yields several parameters allowing for a unique insight into their structure. Moreover, ArGSLab can be used to transform experimental or simulation data into a set of nodes and links for further analysis, and thus provides powerful information for the development of new analyses tailored to user-specific needs. Since the method is developed to analyze any network that is composed of smaller segments, ArGSLab can be applied to networks beyond colloidal gels, such as protein networks,\cite{Becker2011} bacterial aggregates in biofilms\cite{Thomas2018} and biological cell networks.\cite{Hilfiker2019} By providing for an easy means to extract global network parameters, ArGSLab provides a powerful method of extracting mesoscopic structural parameters that can be related to microscopic and macroscopic properties, for instance obtained from light scattering or rheology.

\section{Materials and Methods}
\label{secmatmet}
Dataset 1 is the same as dataset C5 in Ref.~\citenum{Schmiedeberg2016}. Particles are PMMA spheres with a diameter $\sigma = 1.72 \mu$m, aggregated due to depletion forces. We analyze 30 stacks with total imaged volume $8.85\times10^{4} \mu\text{m}^{3}$ or $1.72\times10^4 \sigma^{3}$ per stack, voxel dimensions of 0.061$\times$0.061$\times$0.12 $\sigma^3$, and a volume fraction $\phi$ of approximately $0.2$. Further details can be found in Ref.~\citenum{Schmiedeberg2016}.\\
Dataset 2 consists of polystyrene/poly-N-Isopropylacrylamide core-shell microgels, synthesized using a method described in Ref.~\citenum{Schurtenberger2020}. In collapsed form, $\sigma = 398$nm and particles aggregate irreversibly due to a combination of van der Waals interactions and other short-range attractions. We analyzed 40 image stacks with a total imaged volume per stack of $1.08\times10^{4} \mu\text{m}^{3}$ or $1.71\times10^5 \sigma^{3}$, voxel dimensions of 0.054$\times$0.054$\times$0.12 $\sigma^3$, with $\phi = 0.05$. The samples are prepared and imaged using identical methods as described in Ref.~\citenum{Stenhammar2020}.\\
Dataset 3 was obtained from Brownian dynamics using particles with attractive patches icosahedrally placed on the surface as described in Ref.~\citenum{Stenhammar2020}. We analyze the final structures from four different Brownian dynamics simulations. To compensate for local fluctuations, the results from 10 temporally close coordinate diagrams were averaged for each simulation. The total system volume was $1.0\times10^5 \sigma^{3}$ with $\phi = 0.05$. Further details can be found in Ref.~\citenum{Stenhammar2020}.\\

\section*{Computational Efficiency}
On an Intel\textsuperscript{\textregistered} octacore i7-7700 CPU with 3.60 GHz running Windows 10 and MATLAB 2021a, extracting a skeleton and subsequent analysis from one image from dataset 1 (512x512x151 pixels, $\phi = 0.2$) takes approximately 5 minutes. Starting from the coordinate diagram only adds a few seconds to this calculation time. Additional visualization and optimization visualization adds 6 minutes. The code has been tested on iOS, Windows and Linux platforms, with comparable computation times. Decreasing image size and $\phi$ reduces computational time.

\section*{Resources}
A project page for the package exists at GitHub.com (github.com/jimmink/argslab), where the standalone MATLAB package and GUI for Windows, iOS and Linux are freely available, together with a user manual and example datasets. An idealized, computer-generated image stack is also available, intended to help users familiarize themselves with ArGSLab and its parameters. A walkthrough of the analysis of this image is available as Supplementary Information.

\section*{Acknowledgements}
We gratefully acknowledge financial support from the Alexander von Humboldt Foundation (JNI), the European Research Council (ERC-339678-COMPASS (PS)) and the Swedish Research Council (Grant numbers 2018-04627 (PS) and 2019-03718 (JS)).

\section*{Supplementary Information}
A walkthrough of the ArGSLab analysis of the freely available benchmarking image.

\footnotesize{
\bibliography{bib_cry1}
}

\end{document}